\documentclass[journal]{IEEEtran}

\usepackage{amssymb}
\usepackage{amsmath}
\usepackage{amsfonts}
\usepackage{algorithm}
\usepackage{graphicx}
\usepackage{fancyhdr}

\newtheorem{theorem}{Theorem}

\newtheorem{definition}[theorem]{Definition}

\newtheorem{lemma}[theorem]{Lemma}

\newtheorem{problem}[theorem]{Problem}

\newtheorem{remark}[theorem]{Remark}

\numberwithin{equation}{section}
\numberwithin{equation}{subsection}
\numberwithin{theorem}{subsection}
\input{tcilatex}
\begin{document}

\title{Delay-Doppler Channel Estimation with Almost Linear Complexity}
\author{Alexander Fish, Shamgar Gurevich, Ronny Hadani, Akbar Sayeed, and
Oded Schwartz\thanks{%
A. Fish is with School of Mathematics and Statistics, University of Sydney,
Sydney, NSW 2006, Australia. Email: alexander.fish@sydney.edu.au.}\thanks{%
S. Gurevich is with the Department of Mathematics, University of Wisconsin,
Madison, WI 53706, USA. Email: shamgar@math.wisc.edu.}\thanks{%
R. Hadani is with the Department of Mathematics, University of Texas,
Austin, TX 78712, USA. Email: hadani@math.utexas.edu.}\thanks{%
A. Sayeed is with the Department of Electrical and Computer Engineering,
University of Wisconsin, Madison, WI 53706, USA. Email: akbar@engr.wisc.edu.}%
\thanks{%
O. Schwartz is with the Department of Electrical Engineering and Computer
Science, University of California, Berkeley, CA 94720, USA. Email:
odedsc@eecs.berkeley.edu.}\medskip }
\specialpapernotice{\textit{To Solomon Golomb for the occasion of his 80 birthday mazel tov}}
\maketitle

\begin{abstract}
A fundamental task in wireless communication is \textit{channel estimation}:
Compute the channel parameters a signal undergoes while traveling from a
transmitter to a receiver. In the case of delay-Doppler channel, a widely
used method is the \textit{matched filter} algorithm. It uses a
pseudo-random sequence of length $N,$ and, in case of non-trivial relative
velocity between transmitter and receiver, its computational complexity is $%
O(N^{2}\log N)$. In this paper we introduce a novel approach of designing
sequences that allow faster channel estimation. Using group representation
techniques we construct sequences, which enable us to introduce a new
algorithm, called the \textit{flag method, }that significantly improves the
matched filter algorithm. The flag method finds the channel parameters in $%
O(m\cdot N\log N)$ operations, for channel of sparsity $m$. We discuss
applications of the flag method to GPS, radar system, and mobile
communication as well.
\end{abstract}


\markboth{{\it Delay-Doppler Channel Estimation with Almost Linear Complexity
--- By Fish, Gurevich, Hadani, Sayeed, and Schwartz}}{}

\section{\textbf{Introduction}}
\label{Intro}

\PARstart{A}{ basic} step in many wireless communication protocols \cite{TV}
is \textit{channel estimation}: learning the channel parameters a signal
undergoes while traveling from a transmitter to a receiver. In this paper we
develop an efficient algorithm for delay-Doppler (also called
time-frequency) channel estimation. Our algorithm provides a striking
improvement over current methods in the presence of high relative velocity
between a transmitter and a receiver. The latter scenario occurs in GPS,
radar systems, mobile communication of fast moving users, and very high
frequency (GHz) communication.

Throughout this paper we denote by $\mathcal{H}=%
\mathbb{C}
(%
\mathbb{Z}
_{N})$ the vector space of complex valued functions on the set of integers $%
\mathbb{Z}
_{N}=\{0,1,...,N-1\}$ equipped with addition and multiplication modulo $N.$
We assume that $N$ is an odd prime number. The vector space $\mathcal{H}$ is
endowed with the inner product 
\begin{equation*}
\left\langle f_{1},f_{2}\right\rangle =\tsum\limits_{n\in 
\mathbb{Z}
_{N}}f_{1}[n]\overline{f_{2}[n]},
\end{equation*}%
for $f_{1},f_{2}\in \mathcal{H}$, \ and referred to as the \textit{Hilbert
space of (digital) sequences}.

Let us start with example.

\nointerlineskip\vbox to 0pt { É }\nointerlineskip

\subsection{\textbf{Example: The GPS Problem\label{Ex-GPS} }}

\begin{figure}[h!] 
\includegraphics[clip,height=8cm]{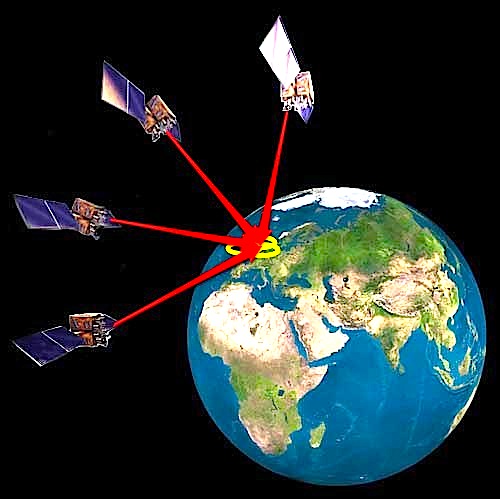}\\
\caption{Satellites communicate location in GPS.}
\label{GPS}
\end{figure}

A client on the earth surface wants to know his/her geographical location.
The Global Positioning System (GPS) was built to fulfill this task. Its
mathematical model works as follows \cite{K}. Satellites send to earth their
location---see Figure \ref{GPS} for illustration.


For simplicity, the location of a
satellite is a bit $b\in \{\pm 1\}.$ The satellite transmits to the earth
its sequence $S\in \mathcal{H}$ of norm one multiplied by its location $b.$
We assume, for simplicity, that the sequence travels through only one path.
Hence, the client receives the sequence $R\in \mathcal{H}$ of the form%
\footnote{%
We denote $i=\sqrt{-1}.$} 
\begin{equation}
R[n]=b\cdot \alpha _{0}\cdot e^{\frac{2\pi i}{N}\omega _{0}\cdot n}\cdot
S[n+\tau _{0}]\text{ }+\mathcal{W}[n],  \label{GPS-P}
\end{equation}%
where $\alpha _{0}\in 
\mathbb{C}
$ is the \textit{complex amplitude,} with $\left\vert \alpha _{0}\right\vert
\leq 1,$ $\omega _{0}\in 
\mathbb{Z}
_{N}$ encodes the radial velocity of the satellite with respect to the
client, $\tau _{0}\in 
\mathbb{Z}
_{N}$ encodes the distance between the satellite and the client\footnote{%
Using $\tau _{0}$ we can compute \cite{K} the distance from the satellite to
the client, assuming a line of sight between them.}, and $\mathcal{W}$ is a
random white noise\footnote{%
In this paper, a random white noise will always be assumed to have mean zero.%
}. The problem of GPS can be formulated as follows:\bigskip

\begin{problem}[\textbf{The GPS Problem}]
\label{GPS-p}Design $S\in \mathcal{H}$, and an effective method of
extracting $(b,\tau _{0})$ from $S$ and $R$ satisfying (\ref{GPS-P}%
).\smallskip \medskip\ 
\end{problem}

In practice, the satellite transmits $S=S_{1}+b\cdot S_{2},$ where $%
S_{1},S_{2}$ are almost orthogonal in some appropriate sense. Then $(\alpha
_{0},\tau _{0},\omega _{0}),$ and $(b\cdot \alpha _{0},\tau _{0},\omega
_{0}) $ are computed using $S_{1},$ and $S_{2},$ respectively, concluding
with the derivation of the bit $b$. A client can compute his/her location by
knowing the locations of at least three satellites and distances to them.
The GPS\ problem is an example of channel estimation task. We would like now
to describe the more general channel estimation problem that we are going to
solve.

\subsection{\textbf{Channel Estimation Problem}}

We consider the following mathematical model of time-frequency channel
estimation \cite{TV}. There exists a collection of users, each one holds a
sequence from $\mathcal{H}$ known to a base station (receiver). The users
transmit their sequences to the base station. Due to multipath effects---see
Figure \ref{multi3} for illustration---the sequences undergo \cite{SA, TV}
several time-frequency shifts as a result of reflections from various
obstacles. We make the standard assumption of almost-orthogonality between
sequences of different users. Hence, if a user transmits $S\in \mathcal{H}$,
then the base station receives $R\in \mathcal{H}$ of the form 
\begin{equation}
R[n]=\sum_{k=1}^{m}\alpha _{k}\cdot e^{\frac{2\pi i}{N}\omega _{k}\cdot
n}\cdot S[n+\tau _{k}]+\mathcal{W}[n],\text{ \ }n\in 
\mathbb{Z}
_{N},  \label{RCE}
\end{equation}%
where $m$ denotes the number of paths the transmitted sequence traveled, $%
\alpha _{k}\in 
\mathbb{C}
$ is the \textit{complex multipath amplitude} along path $k$\textit{,} with $%
\sum_{k=1}^{m}\left\vert \alpha _{k}\right\vert ^{2}\leq 1,$ $\omega _{k}\in 
\mathbb{Z}
_{N}$ depends on the relative velocity along path $k$ of the transmitter
with respect to a base station, $\tau _{k}\in 
\mathbb{Z}
_{N}$ encodes the delay along path $k,$ and $\mathcal{W}\in \mathcal{H}$
denotes a random white noise$.$ The parameter $m$ will be called the \textit{%
sparsity} of the channel.\medskip\ The objective is:\smallskip\ \medskip

\begin{problem}[\textbf{The Channel Estimatiom Problem}]
\label{CE}Design $S\in \mathcal{H}$, and an effective method of extracting
the channel parameters $(\alpha _{k},\tau _{k},\omega _{k})$, $\ k=1,...,m,$
from $S$ and $R$ satisfying (\ref{RCE}).\smallskip \medskip


\begin{figure}[h!] 
\includegraphics[clip,height=6cm]{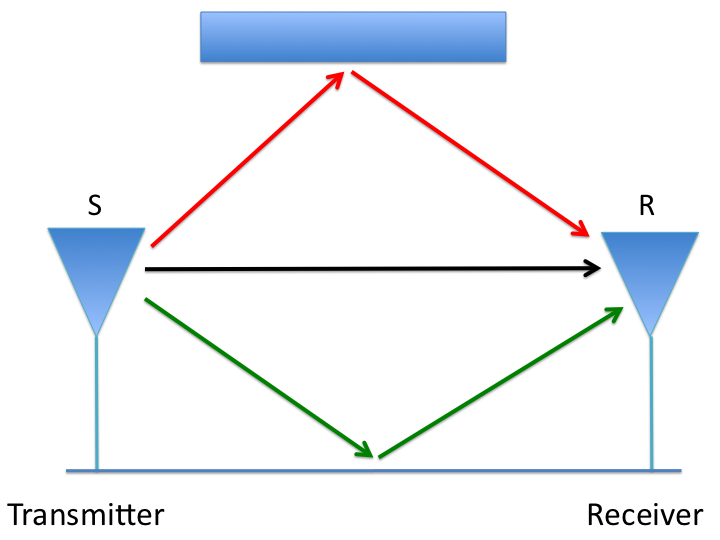}\\
\caption{Three paths scenario.}
\label{multi3}
\end{figure}

\end{problem}

To suggest a solution to Problem \ref{CE}, we start with a simpler
variant.\medskip

\subsection{\textbf{The Time-Frequency Shift (TFS) Problem}}

Suppose the transmitter and the receiver sequences $S,R\in \mathcal{H}$ are
related by 
\begin{equation}
R[n]=e^{\frac{2\pi i}{N}\omega _{0}\cdot n}\cdot S[n+\tau _{0}]\text{ }+%
\mathcal{W}[n],\text{ \ \ }n\in 
\mathbb{Z}
_{N},  \label{RTFS}
\end{equation}%
where $\mathcal{W}\in \mathcal{H}$ denotes a random white noise, and $(\tau
_{0},\omega _{0})\in 
\mathbb{Z}
_{N}\times 
\mathbb{Z}
_{N}$. The pair $(\tau _{0},\omega _{0})$ is called the \textit{%
time-frequency shift}, and the vector space $V=%
\mathbb{Z}
_{N}\times 
\mathbb{Z}
_{N}$ is called the\ \textit{time-frequency plane}. We would like to solve
the following:\smallskip \medskip

\begin{problem}[\textbf{Time-Frequency Shift (TFS)}]
\label{TFS}Design $S\in \mathcal{H}$, and an effective method of extracting
the time-frequency shift $(\tau _{0},\omega _{0})$ from $S$ and $R$
satisfying (\ref{RTFS}).\smallskip \medskip
\end{problem}

\subsection{\textbf{The Matched Filter (MF) Algorithm}}

A classical solution \cite{GG,GHS1,GHS2,HCM,TV,V,WG} to Problem \ref{TFS},
is the \textit{matched filter algorithm}. We define the following matched
filter (MF) matrix of $R$ and $S$:%
\begin{equation*}
\mathcal{M}(R,S)[\tau ,\omega ]=\left\langle R[n],e^{\frac{2\pi i}{N}\omega
\cdot n}\cdot S[n+\tau ]\right\rangle \text{, \ \ }(\tau ,\omega )\in V.
\end{equation*}%
A direct verification shows that for $\zeta _{0}=e^{\frac{2\pi i}{N}(\tau
\omega _{0}-\omega \tau _{0})}$, with probability one, we have 
\begin{eqnarray*}
\mathcal{M}(R,S)[\tau ,\omega ] &=&\zeta _{0}\cdot \mathcal{M}(S,S)[\tau
-\tau _{0},\omega -\omega _{0}] \\
&&+O(\frac{NSR}{\sqrt{N}}),
\end{eqnarray*}%
where $NSR\approx \frac{1}{SNR}$, i.e., essentially\footnote{%
The precise relation is $NSR=\frac{\sqrt{2\log \log N}}{SNR}$ by the \textit{%
law of the iterated logarithm.}} the inverse of the signal-to-noise ratio
between the sequences $S$ and $\mathcal{W}.$ For simplicity, we assume, for
the rest of the paper, that the $NSR$ is of size $O(1).$ In order to extract
the time-frequency shift $(\tau _{0},\omega _{0}),$ it is \textquotedblleft
standard"\footnote{%
For example in spread-spectrum communication systems.} (see \cite%
{GG,GHS1,GHS2,HCM,TV,V,WG}) to use pseudo-random sequence $S\in \mathcal{H}$
of norm one. In this case $\mathcal{M}(S,S)[\tau -\tau _{0},\omega -\omega
_{0}]=1$ for $(\tau ,\omega )=(\tau _{0},\omega _{0})$, and of order $O(%
\frac{1}{\sqrt{N}})$ if $(\tau ,\omega )\neq (\tau _{0},\omega _{0})$. Hence,%
\begin{equation}
\mathcal{M}(R,S)[\tau ,\omega ]=\left\{ 
\begin{array}{c}
1+\varepsilon _{N}\text{, if }(\tau ,\omega )=(\tau _{0},\omega _{0}); \\ 
\varepsilon _{N}\text{, \ \ \ \ \ if }(\tau ,\omega )\neq (\tau _{0},\omega
_{0}),%
\end{array}%
\right.  \label{Peak}
\end{equation}%
where $\varepsilon _{N}=O(\frac{1}{\sqrt{N}}).$\bigskip


\begin{figure}[h!] 
\includegraphics[clip,height=5cm]{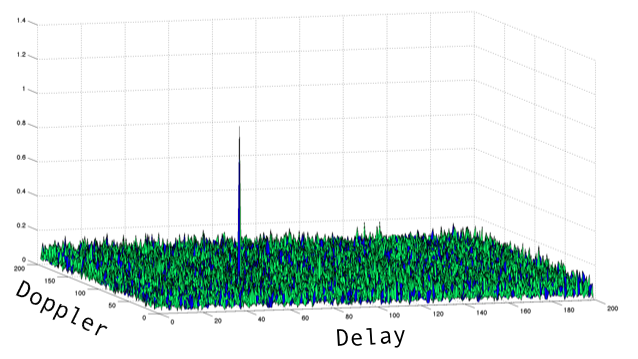}\\
\caption{$\left\vert \mathcal{M}(R,S)\right\vert $
with pseudo-random $S$, and $(\protect\tau _{0},\protect\omega %
_{0})=(50,50).$}
\label{Spike}
\end{figure}

Identity (\ref{Peak})---see Figure \ref{Spike} for a
demonstration---suggests the following \textquotedblleft entry-by-entry"
solution to TFS\ problem: Compute the matrix $\mathcal{M}(R,S),$ and choose $%
(\tau _{0},\omega _{0})$ for which $\mathcal{M}(R,S)[\tau _{0},\omega
_{0}]\approx 1.$ However, this solution of TFS problem is significantly
expensive in terms of arithmetic complexity, i.e., the number of
multiplication and addition operations is $O(N^{3}).$ One can do better
using a \textquotedblleft line-by-line" computation. This is due to the next
observation.\smallskip

\begin{remark}[\textbf{FFT}]
\label{FFT}The restriction of the matrix $\mathcal{M(}R,S)$ to any line (not
necessarily through the origin) in the time-frequency plane $V,$ is a
convolution that can be computed, using the fast Fourier transform (FFT), in 
$O(N\log N)$ operations.\smallskip\ For details see Section \ref{MF on Line}.
\end{remark}

As a consequence of Remark \ref{FFT}, one can solve TFS\ problem in $%
O(N^{2}\log N)$ operations.\medskip

\subsection{\textbf{The Fast Matched Filter (FMF) Problem}}

To the best of our knowledge, the \textquotedblleft line-by-line"
computation is also the fastest known method \cite{OMB}. If $N$ is large
this may not suffice. For example in applications to GPS \cite{Ag}, as in
Problem \ref{GPS-p} above, we have $N\geq 1000.$ This leads to the
following:\smallskip

\begin{problem}[\textbf{The Fast Matched Filter Problem}]
\label{FMF}Solve the TFS problem in almost linear complexity.\smallskip
\medskip
\end{problem}

Note that computing one entry in $\mathcal{M(}R,S)$ already takes $O(N)$
operations.

\subsection{\textbf{The Flag Method\label{FM}}}

\begin{figure}[h!] 
\includegraphics[clip,height=5cm]{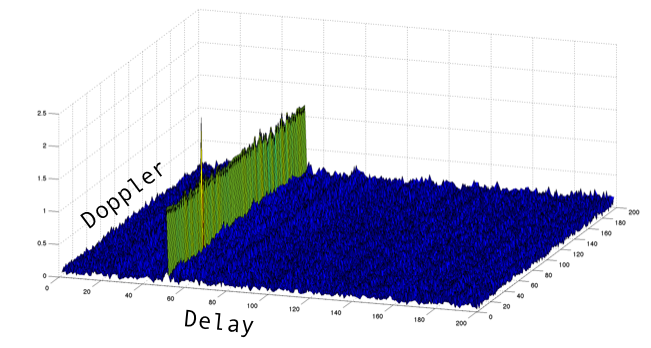}\\
\caption{$\left\vert \mathcal{M}(R,S_{L})\right\vert $ for a
flag $S_{L}$ with $L=\{(0,\protect\omega )\},$ and $(\protect%
\tau _{0},\protect\omega _{0})=(50,50).$}
\label{Degel}
\end{figure}

In this paper we introduce the \textit{flag method} to propose a solution to
FMF problem. The idea is, first, to find a line on which the time-frequency
shift is located, and, then, to search on the line to find the
time-frequency shift. We associate with the $N+1$ lines $L_{j},$ $%
j=1,...,N+1,$ through the origin in $V,$ a system of \textquotedblleft
almost orthogonal" sequences $S_{L_{j}}\in \mathcal{H},$ that we call 
\textit{flags}. They satisfy---see Figure \ref{Degel} for illustration---the
following \textquotedblleft flag property"\footnote{%
In linear algebra, a pair $(\ell _{0},L)$ consisting of a line $L\subset V,$
and a point $\ell _{0}\in L$, is called a \textit{flag}.}: For a sequence $R$
given by (\ref{RTFS}) with $S=S_{L_{j}},$ we have 
\begin{eqnarray}
&&\mathcal{M}(R,S_{L_{j}})[\tau ,\omega ]  \label{flag} \\
&=&\left\{ 
\begin{array}{c}
2+\varepsilon _{N}\text{,\ \ \ if }(\tau ,\omega )=(\tau _{0},\omega _{0});%
\text{ \ \ \ \ \ \ \ \ \ \ \ \ \ } \\ 
1+\varepsilon _{N}\text{, \ \ in }\left\vert \cdot \right\vert \text{ if }%
(\tau ,\omega )\in L_{j}^{\prime }\smallsetminus (\tau _{0},\omega _{0}); \\ 
\text{ }\varepsilon _{N}\text{,\ \ \ \ \ \ \ \ if }(\tau ,\omega )\in
V\smallsetminus L_{j}^{\prime },\text{\ \ \ \ \ \ \ \ \ \ \ \ \ \ \ }%
\end{array}%
\right.  \notag
\end{eqnarray}%
where $\varepsilon _{N}=O(\frac{1}{\sqrt{N}}),$ $\left\vert \cdot
\right\vert $ denotes absolute value, and $L_{j}^{\prime }$ is the shifted
line $L_{j}+(\tau _{0},\omega _{0})$. The \textquotedblleft almost
orthogonality" of sequences means $\left\vert \mathcal{M}%
(S_{L_{i}},S_{L_{j}})[\tau ,\omega ]\right\vert =O(\frac{1}{\sqrt{N}}),$ for
every $(\tau ,\omega )$, $i\neq j.$\ \medskip


\medskip

In addition, for $S_{L}$ and $R$ satisfying (\ref{flag}), we have the
following search method to solve FMF\ problem: \smallskip \smallskip

\underline{\textbf{Flag Algorithm}}\medskip

\begin{description}
\item[\textbf{Step 1.}] \ \ Choose a line $L^{\bot }$ transversal to $%
L.\smallskip $

\item[\textbf{Step 2.}] \ \ Compute $\mathcal{M}(R,S_{L})$ on $L^{\bot }$.
Find $(\tau ,\omega )$ such that $\left\vert \mathcal{M}(R,S_{L})[\tau
,\omega ]\right\vert $ $\approx 1$, i.e., $(\tau ,\omega )$ on the shifted
line $L+(\tau _{0},\omega _{0}).\smallskip $

\item[\textbf{Step 3.}] \ \ Compute $\mathcal{M}(R,S_{L})$ on $L+(\tau
_{0},\omega _{0})$ and find $(\tau ,\omega )$ such that $\left\vert \mathcal{%
M}(R,S_{L})[\tau ,\omega ]\right\vert \approx 2.\smallskip $
\end{description}

\begin{figure}[h!] 
\includegraphics[clip,height=5cm]{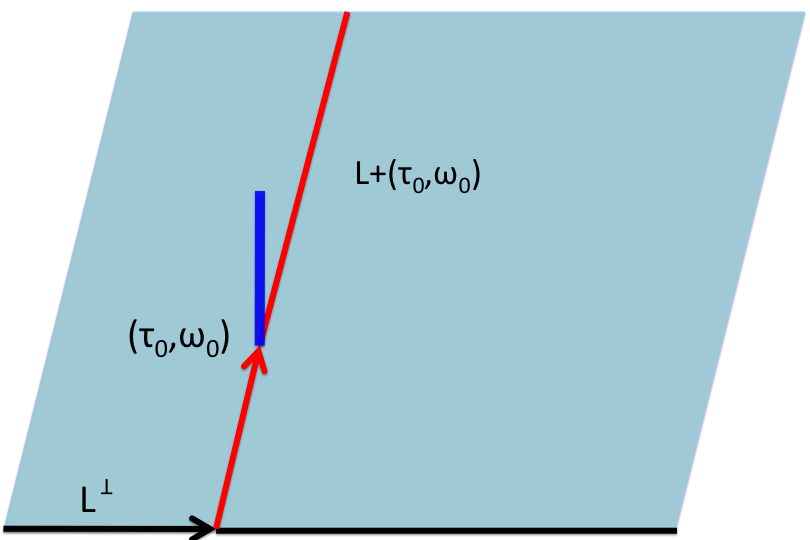}\\
\caption{Diagram of flag algorithm.}
\label{flag-algorithm}
\end{figure}

The complexity of the flag algorithm---see Figure \ref{flag-algorithm} for a
demonstration---is $O(N\log N),$ using the FFT.


This completes our solution of Problem \ref{FMF}---The Fast Matched Filter
Problem.

\subsection{\textbf{Solution to the GPS and Channel Estimation Problems\label%
{SCE}}}

Let $L\subset V$ be a line through the origin.
\medskip

\begin{definition}[\textbf{Genericity}]
We say that the points $(\tau _{k},\omega _{k})\in V,$ $k=1,...,m,$ are $L$-%
\textit{generic} if no two of them lie on a shift of $L$, i.e., on $L+v$,
for some $v\in V.\medskip $
\end{definition}

Looking back to Problem \ref{CE}, we see that, under genericity assumptions,
the flag method provides a fast computation, in $O(m\cdot N\log N)$
operations, of the channel parameters of channel with sparsity $m$ In
particular, it calculates the GPS parameters---see Problem \ref{GPS-P}---in $%
O(N\log N)$ operations. Indeed, Identity (\ref{flag}), together with the
almost orthogonality between flag sequences, implies that 
\begin{equation*}
\alpha _{k}\approx \mathcal{M}(R,S_{L})[\tau _{k},\omega _{k}]/2,\text{ \ \ }%
k=1,...,m,
\end{equation*}%
where $R$ is the sequence (\ref{RCE}), with $S=S_{L},$ assuming that and \ $%
(\tau _{k},\omega _{k})$'s are $\ L$-generic.$\ $So we can adjust the flag
algorithm as follows:

\begin{itemize}
\item Compute $\mathcal{M}(R,S_{L})$ on $L^{\bot }$. Find all $(\tau ,\omega
)$'s such that $\left\vert \mathcal{M}(R,S_{L})[\tau ,\omega ]\right\vert $
is sufficiently large, i.e., find all the shifted lines $L+(\tau _{k},\omega
_{k})$'s$.\smallskip $

\item Compute $\mathcal{M}(R,S_{L})$ on each line $L+(\tau _{k},\omega
_{k}), $ and find $(\tau ,\omega )$ such that $\left\vert \mathcal{M}%
(R,S_{L})[\tau ,\omega ]\right\vert $ is maximal on that line, i.e., $(\tau
,\omega )=(\tau _{k},\omega _{k})$ and $\alpha _{k}\approx \mathcal{M}%
(R,S_{L})[\tau _{k},\omega _{k}]/2.$
\end{itemize}

\begin{figure}[h!] 
\includegraphics[clip,height=5cm]{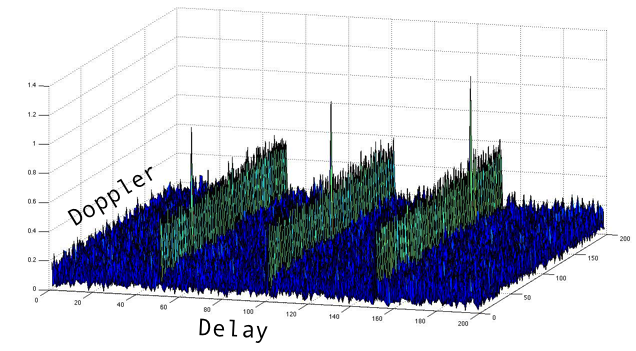}\\
\caption{$%
\left\vert \mathcal{M}(R,S_{L})\right\vert ,$ for $L=\{(0,\protect\omega )\},
$ and $(\protect\alpha _{k},\protect\tau _{k},\protect\omega _{k})=(\frac{1}{%
\protect\sqrt{3}},50k,50k),$ $k=1,2,3.$}
\label{channel}
\end{figure}

Figure \ref{channel} provides a visual illustration for the matched filter
matrix in three paths scenario.


This completes our solutions of Problem \ref{CE}---The Channel Estimation
Problem, and of Problem \ref{GPS-P}---The GPS Problem.

\subsection{\textbf{Applications to Radar and Mobile Communication}}

The flag method provides a significant improvement over the current channel
estimation algorithms in the presence of high velocities. The latter occurs
in systems such as GPS, radar, and mobile communication of fast moving
users. In Subsection \ref{Ex-GPS}, we described the GPS problem, and in
Subsection \ref{SCE} its effective solution using the flag method. It is
easy to see that the flag method suggests a solution to the GPS problem also
in the multipath scenario. In this section we demonstrate application of the
flag method to radar, and mobile communication.\medskip

\subsubsection{\textbf{\ Application to Radar}}

\begin{figure}[h!] 
\includegraphics[clip,height=5cm]{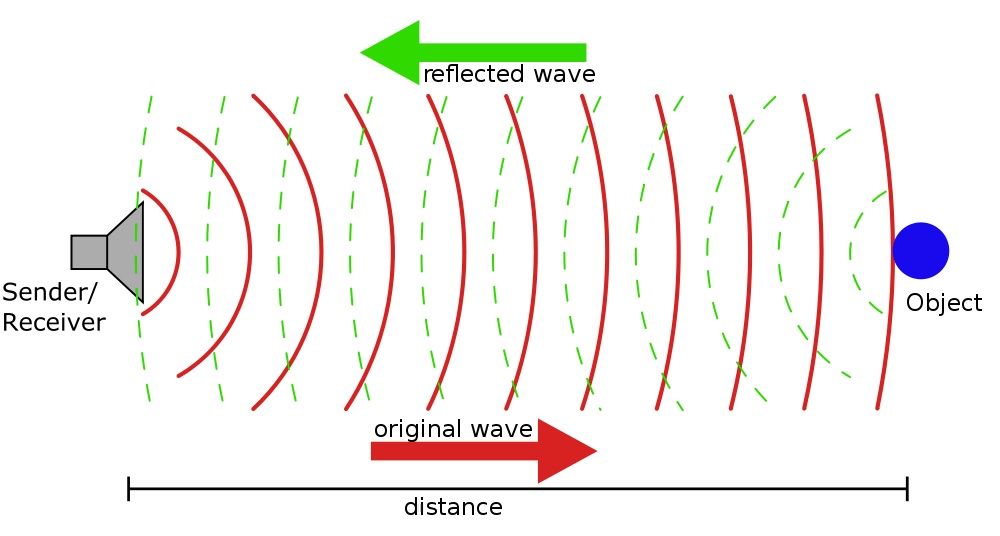}\\
\caption{Radar transmits wave and recieves echo.}
\label{radar}
\end{figure}

The model of radar works as follows \cite{HCM}. A radar transmits---Figure %
\ref{radar} illustrates the case of one target---a sequence $S\in \mathcal{H}
$ which bounces back from $m$ targets. The sequence $R\in \mathcal{H}$ which
is received as an echo has the form%
\begin{equation*}
R[n]=\sum_{k=1}^{m}\alpha _{k}\cdot e^{\frac{2\pi i}{N}\omega _{k}\cdot
n}\cdot S[n+\tau _{k}]+\mathcal{W}[n],\ n\in 
\mathbb{Z}
_{N},
\end{equation*}%
where $\alpha _{k}\in 
\mathbb{C}
$ is the complex multipath amplitude along path $k$\textit{,} with $%
\sum_{k=1}^{m}\left\vert \alpha _{k}\right\vert ^{2}\leq 1,$ $\omega _{k}\in 
$ $%
\mathbb{Z}
_{N}$ encodes the radial velocity of target $k$ with respect to the radar, $%
\tau _{k}\in 
\mathbb{Z}
_{N}$ encodes the distance between target $k$ and the radar, and $\mathcal{W}
$ is a random white noise.\smallskip \smallskip\ 

In order to determine the location of the targets we need to solve \cite{L}
the following.

\begin{problem}[\textbf{The Radar Problem}]
\medskip \label{RP}Having $R$ and $S$, compute the parameters $(\tau
_{k},\omega _{k}),$ $k=1,...,m.\smallskip \smallskip $
\end{problem}

This is essentially the channel estimation problem. Under the genericity
assumption, the flag method solves it in $O(m\cdot N\log N)$ operations.
This completes our solution to Problem \ref{RP}---The Radar Problem.


\ \medskip

\subsubsection{\textbf{Application to Mobile Communication}}

The model of mobile communication works as follows \cite{TV}. A user wants
to deliver a bit of information $b\in \{\pm 1\}$ to a base station. The base
station assigns a sequence $S\in \mathcal{H}$ to the user, and the user
transmits to the base station the sequence $b\cdot S$. The sequence $R\in 
\mathcal{H}$ which is received by the base station is of the form 
\begin{equation*}
R[n]=b\cdot \sum_{k=1}^{m}\alpha _{k}\cdot e^{\frac{2\pi i}{N}\omega
_{k}\cdot n}\cdot S[n+\tau _{k}]+\mathcal{W}[n],\text{ \ }n\in 
\mathbb{Z}
_{N},
\end{equation*}%
where $m$ denotes the number of paths the transmitted sequence traveled, $%
\alpha _{k}\in 
\mathbb{C}
$ is the multipath amplitude along path $k$\textit{,} with $%
\sum_{k=1}^{m}\left\vert \alpha _{k}\right\vert ^{2}\leq 1,$ $\omega _{k}\in 
\mathbb{Z}
_{N}$ depends on the relative velocity along path $k$ of the user with
respect to the base station, $\tau _{k}\in 
\mathbb{Z}
_{N}$ is the delay along path $k,$ and $\mathcal{W}\in \mathcal{H}$ denotes
a random white noise$.$ The main task at the base station is\medskip\ the
following:

\begin{problem}[\textbf{The Mobile Communication Problem}]
\label{MCP}Having $R$ and $S$, compute the bit $b.\medskip $
\end{problem}

\medskip In practice, first the user sends $S$, and the channel estimation
is done. Then the bit $b$ is communicated by sending $b\cdot S$. Finally,
knowing the channel parameters $(\alpha _{k},\tau _{k},\omega _{k}),$ $%
k=1,...,m,\smallskip \smallskip $ the bit is extracted using the formula%
\footnote{%
It is analogous to data modulation using a delay-Doppler rake receiver in
spread-spectrum systems \cite{TV}.}%
\begin{equation*}
b\cdot \sum_{k=1}^{m}|\alpha _{k}|^{2}\approx \langle R,\sum_{k=1}^{m}\alpha
_{k}\cdot e^{\frac{2\pi i}{N}\omega _{k}\cdot n}\cdot S[n+\tau _{k}]\rangle .
\end{equation*}%
The main computational step is the channel estimation which is done by flag
method in $O(m\cdot N\log N)$ operations. This completes our solution to
Problem \ref{MCP}---The Mobile Communication Problem.

\subsection{\textbf{What you can find in this paper}}

\begin{itemize}
\item \textit{In Section \ref{Intro}:} You can read about the flag method for
effective delay-Doppler channel estimation. In addition, concrete
applications to GPS, radar, and mobile communication are discussed.

\item \textit{In Section \ref{H--W Ops}:} You can find the definition and
explicit formulas for the Heisenberg and Weil operators. These operators are
our basic tool in the development of the flag method, in general, and the
flag sequences, in particular.

\item \textit{In Section \ref{H--W Fls}:} You can see the design of the
Heisenberg--Weil flag sequences, using the Heisenberg--Weil operators, and
diagonalization techniques of commuting operators. In addition, the
investigation of the correlation properties of the flag sequences is done in
this section. These properties are formulated in Theorem \ref{HW}, which
guarantees applicability of the Heisenberg-Weil sequences to the flag method.

\item \textit{In Section \ref{H--W For}: }You can get explicit formulas for
large collection of the Heisenberg--Weil flag sequences. In particular,
these formulas enable to generate the sequences using low complexity
algorithm.

\item \textit{In Section \ref{MF on Line}:} You can find the formulas that
suggest fast computation of the matched filter matrix on any line in the
time-frequency plane. These formulas are of crucial importance for the
effectiveness of the flag method.

\item \textit{In Section \ref{Proofs}: }You can find needed proofs and
justifications for all the claims and formulas that appear in the body of
the paper. \medskip
\end{itemize}

\textbf{Acknowledgement. }Warm thanks to J. Bernstein for his support and
encouragement in interdisciplinary research. We are grateful to A. Sahai,
for sharing with us his thoughts, and ideas on many aspects of signal
processing and wireless communication. The project described in this paper
was initiated by a question of M. Goresky and A. Klapper during the
conference SETA2008, we thank them very much. We appreciate the support and
encouragement of N. Boston, R. Calderbank, S. Golomb, G. Gong, O. Holtz, R.
Howe, P. Sarnak, N. Sochen, D. Tse, and A. Weinstein. The research reported
in this paper was partially supported by NSF Grants DMS-1101660, and
DMS-1101698.

\section{\textbf{The Heisenberg and Weil Operators\label{H--W Ops}}}

The flag sequences (see Subsection \ref{FM}) are defined, constructed and
analyzed using two special classes of operators that act on the Hilbert
space of digital sequences. The first class consists of the Heisenberg
operators and is a generalization of the time-shift and frequency-shift
operators. The second class consists of the Weil operators and is a
generalization of the discrete Fourier transform. In this section we recall
the definitions and explicit formulas of these operators.

\subsection{\textbf{The Heisenberg Operators}}

The Heisenberg operators are the unitary transformations that act on the
Hilbert space of digital sequences by 
\begin{equation}
\left\{ 
\begin{array}{c}
\pi (\tau ,\omega ):\mathcal{H\rightarrow \mathcal{H}},\text{ \ }\tau
,\omega \in 
\mathbb{Z}
_{N}; \\ 
\lbrack \pi (\tau ,\omega )f][n]=e^{\frac{2\pi i}{N}\omega \cdot n}\cdot
f[n+\tau ],%
\end{array}%
\right.  \label{Heis}
\end{equation}%
for every $f\in \mathcal{H}$, $n\in 
\mathbb{Z}
_{N}.$

\subsection{\textbf{The Weil Operators}}

Consider the discrete Fourier transform%
\begin{equation*}
\left\{ 
\begin{array}{c}
DFT:\mathcal{H\rightarrow \mathcal{H}}, \\ 
\lbrack DFT(f)][\omega ]=\frac{1}{\sqrt{N}}\tsum\limits_{n=0}^{N-1}e^{-\frac{%
2\pi i}{N}\omega \cdot n}\cdot f[n],%
\end{array}%
\right.
\end{equation*}%
for every $f\in \mathcal{H}$, $\omega \in 
\mathbb{Z}
_{N}.$ It is easy to check that $DFT$ satisfies the following $N^{2}$
identities: 
\begin{equation}
\text{ }DFT\circ \pi (\tau ,\omega )=\pi (-\omega ,\tau )\circ DFT,\text{\ \ 
}\tau ,\omega \in 
\mathbb{Z}
_{N},  \label{SDFT}
\end{equation}%
where $\pi (\tau ,\omega )$ are the Heisenberg operators, and $\circ $
denotes composition of transformations. A version of the celebrated
Stone--von Neumann (S--vN) theorem implies that up to scalar multiple the $%
DFT$ is the unique operator that satisfies (\ref{SDFT}). This means that (%
\ref{SDFT}) is a characterization of the $DFT.$ In \cite{W} Weil generalized
this method and defined many other operators that act on $\mathcal{H}$.
Consider the following collection of matrices 
\begin{equation*}
SL_{2}(%
\mathbb{Z}
_{N})=\left\{ 
\begin{pmatrix}
a & b \\ 
c & d%
\end{pmatrix}%
;\text{ }a,b,c,d\in 
\mathbb{Z}
_{N},\text{ and }ad-bc=1\right\} .
\end{equation*}%
Note that $G=SL_{2}(%
\mathbb{Z}
_{N})$ is a \textit{group} \cite{Ar} with respect to the operation of matrix
multiplication. It is called the \textit{special linear group} of order two
over $%
\mathbb{Z}
_{N}.$ Each element 
\begin{equation*}
g=%
\begin{pmatrix}
a & b \\ 
c & d%
\end{pmatrix}%
\in G,
\end{equation*}%
acts on the time-frequency plane $V=%
\mathbb{Z}
_{N}\times 
\mathbb{Z}
_{N}$ via the change of coordinates 
\begin{equation*}
(\tau ,\omega )\mapsto g\cdot (\tau ,\omega )=(a\tau +b\omega ,c\tau
+d\omega ).
\end{equation*}%
For $g\in G$, let $\rho (g)$ be a linear operator on $\mathcal{H}$ which is
a solution of the following system of $N^{2}$ linear equations:%
\begin{equation}
\Sigma _{g}:\text{ }\rho (g)\circ \pi (\tau ,\omega )=\pi (g\cdot (\tau
,\omega ))\circ \rho (g),\text{\ \ }\tau ,\omega \in 
\mathbb{Z}
_{N},  \label{S}
\end{equation}%
Denote by $\mathrm{Sol}(\Sigma _{g})$ the space of all solutions to System (%
\ref{S}). For example for 
\begin{equation*}
\mathrm{w}=%
\begin{pmatrix}
0 & -1 \\ 
1 & \text{ \ }0%
\end{pmatrix}%
,
\end{equation*}%
which is called the \textit{Weyl }element, we have by (\ref{SDFT}) that $%
DFT\in \mathrm{Sol}(\Sigma _{\mathrm{w}}).$ The general version of the S--vN
theorem implies that $\dim \mathrm{Sol}(\Sigma _{g})=1$, for every $g\in G$.
In fact there exists a special set of solutions. This is the content of the
following result \cite{W}:\smallskip \medskip

\begin{theorem}[\textbf{Weil operators}]
\label{S-vN}There exists a unique collection of solutions $\{\rho (g)\in 
\mathrm{Sol}(\Sigma _{g});$ $\ g\in G\},$ which are unitary operators, and
satisfy the homomorphism condition $\rho (g\cdot h)=\rho (g)\circ \rho (h),$
for every $g,h\in G.\medskip $
\end{theorem}

Denote by $U(\mathcal{H)}$ the collection of all unitary operators on the
Hilbert space $\mathcal{H}$ of digital sequences. Theorem \ref{S-vN}
establishes the map 
\begin{equation}
\rho :G\rightarrow U(\mathcal{H}),  \label{WO}
\end{equation}%
which is called the \textit{Weil representation }\cite{W}\textit{. }We will
call each $\rho (g),$ $g\in G,$ a \textit{Weil operator.\medskip }

\subsubsection{\textbf{Formulas for Weil Operators}}

It will be important for our study to have the following \cite{Ge, GHH}
explicit formulas for the Weil operators:

\begin{itemize}
\item \textit{Fourier. }We have 
\begin{equation}
\left[ \rho 
\begin{pmatrix}
0 & -1 \\ 
1 & \text{ \ }0%
\end{pmatrix}%
f\right] [n]=i^{\frac{N-1}{2}}\cdot DFT(f)[n];  \label{F}
\end{equation}

\item \textit{Chirp. }We have%
\begin{equation}
\left[ \rho 
\begin{pmatrix}
1 & 0 \\ 
c & 1%
\end{pmatrix}%
f\right] [n]=e^{\frac{2\pi i}{N}(-2^{-1}cn^{2})}\cdot f[n];  \label{C}
\end{equation}

\item \textit{Scaling. We have}
\end{itemize}

\begin{equation}
\left[ \rho 
\begin{pmatrix}
a & 0 \\ 
0 & a^{-1}%
\end{pmatrix}%
f\right] [n]=\QOVERD( ) {a}{N}f[a^{-1}n],  \label{Sc}
\end{equation}%
for every $f\in \mathcal{H}$, $\ 0\neq a,c,n\in 
\mathbb{Z}
_{N},$ where $\QOVERD( ) {a}{N}$ is the \textit{Legendre symbol }which is
equal to $1$ if $a$ is a square modulo $N,$ and $-1$ otherwise, and in (\ref%
{C}) we denote $2^{-1}=\frac{N+1}{2}$ the inverse of $2$ modulo $N.$ \textit{%
\ }

The group $G$ admits \cite{B} the \textit{Bruhat decomposition }%
\begin{equation*}
G=UA\cup U\mathrm{w}UA,
\end{equation*}%
where $U\subset G$ denotes the \textit{unipotent }subgroup\textit{\ }%
\begin{equation*}
U=\left\{ 
\begin{pmatrix}
1 & 0 \\ 
c & 1%
\end{pmatrix}%
;\text{ \ }c\in 
\mathbb{Z}
_{N}\right\} ,
\end{equation*}%
and $A\subset G$ denotes the \textit{diagonal }subgroup%
\begin{equation}
A=\left\{ 
\begin{pmatrix}
a & 0 \\ 
0 & a^{-1}%
\end{pmatrix}%
;\text{ \ }0\neq a\in 
\mathbb{Z}
_{N}\right\} .  \label{D}
\end{equation}%
This means that every element $g\in G$ can be written in the form 
\begin{equation*}
g=u\cdot s\text{ \ or \ }g=u^{\prime }\cdot \mathrm{w}\cdot u^{\prime \prime
}\cdot s^{\prime }\text{ }
\end{equation*}%
where $u,u^{\prime },u^{\prime \prime }\in U,$ $s,s^{\prime }\in A,$ and $%
\mathrm{w}$ is the Weyl element. Hence, because $\rho $ is homomorphism,
i.e., $\rho (g\cdot h)=\rho (g)\circ \rho (h)$ for every $g,h\in G,$ we
deduce that formulas (\ref{F}), (\ref{C}), and (\ref{Sc}), extend to
describe all the Weil operators.

\section{\textbf{Sequence Design: Heisenberg--Weil Flags\label{H--W Fls}}}

The flag sequences, that play the main role in the flag method, are of a
special type. \ Each of them is a sum of a pseudorandom sequence and a
structural sequence. The first has the MF matrix which is almost delta
function at the origin, and the MF matrix of the second is supported on a
line. The design of these sequences is done using group representation
theory. The pseudorandom sequences are designed \cite{GHS1, GHS2, WG} using
the Weil representation operators (\ref{WO}), and will be called \textit{%
Weil (spike) sequences}\footnote{%
For the purpose of the Flag method, other pseudorandom signals may work.}.
The structural sequences are designed \cite{H, HCM} using the Heisenberg
representation operators (\ref{Heis}), and will be called \textit{Heisenberg
(line) sequences.} We call the collection of all flag sequences, the
Heisenberg--Weil flag system. In this section we study constructions, and
properties of these sequences.\medskip

\subsection{\textbf{The Heisenberg (Lines) System\label{HS}}}

The operators (\ref{Heis}) obey the Heisenberg commutation relations$\frac{{}%
}{{}}$ 
\begin{equation*}
\pi (\tau ,\omega )\circ \pi (\tau ^{\prime },\omega ^{\prime })=e^{\frac{%
2\pi i}{N}(\tau \omega ^{\prime }-\omega \tau ^{\prime })}\cdot \pi (\tau
^{\prime },\omega ^{\prime })\circ \pi (\tau ,\omega ).
\end{equation*}%
The expression $\tau \omega ^{\prime }-\omega \tau ^{\prime }$ vanishes if $%
(\tau ,\omega )$, $(\tau ^{\prime },\omega ^{\prime })$ are on the same line
through the origin. Hence, for a given line $L\subset V=%
\mathbb{Z}
_{N}\times 
\mathbb{Z}
_{N}$ we have a commutative collection of unitary operators%
\begin{equation}
\pi (\ell ):\mathcal{H\rightarrow \mathcal{H}},\text{ }\ell \in L.
\label{piL}
\end{equation}%
Explicit version of simultaneous diagonalization theorem from linear algebra
implies \cite{H, HCM} the existence of a natural orthonormal basis $\mathcal{%
B}_{L\text{ }}$for $\mathcal{H}$ consisting of common eigensequences for all
the operators (\ref{piL}) 
\begin{equation*}
\left\{ 
\begin{array}{c}
\mathcal{B}_{L\text{ }}=\{f_{L_{\psi }}\}; \\ 
\pi (\ell )f_{L_{\psi }}=\psi (\ell )f_{L_{\psi }},\text{ \ }\ell \in L,%
\text{ }%
\end{array}%
\right.
\end{equation*}%
where $\psi $ runs over \textit{characters} of $L$, i.e., functions $\psi
:L\rightarrow 
\mathbb{C}
^{\ast }=%
\mathbb{C}
-0$, with $\psi (\ell +\ell ^{\prime })=\psi (\ell )\psi (\ell ^{\prime })$,
for every $\ell ,\ell ^{\prime }$ $\in L.$ The system of all such bases $%
\mathcal{B}_{L},$ where $L$ runs over all lines through the origin in $V,$
will be called the \textit{Heisenberg (lines) system}. We use the following
result \cite{H, HCM}:\smallskip \medskip

\begin{theorem}
\label{HT}The Heisenberg system satisfies the following properties:
\end{theorem}

\begin{enumerate}
\item \textit{Line. }For every line $L\subset V$, and every $f_{L}\in 
\mathcal{B}_{L}$, we have 
\begin{equation*}
\mathcal{M}(f_{L},f_{L})[\tau ,\omega ]=\left\{ 
\begin{array}{c}
1,\text{ if }\left( \tau ,\omega \right) =(0,0);\text{ \ \ \ \ \ \ \ \ \ \ \
\ \ \ } \\ 
1\text{, \ in }\left\vert \cdot \right\vert \text{ if }\left( \tau ,\omega
\right) \in L\smallsetminus (0,0);\text{ \ } \\ 
0,\text{\ if }\left( \tau ,\omega \right) \notin L.\text{ \ \ \ \ \ \ \ \ \
\ \ \ \ \ \ \ \ \ \ }%
\end{array}%
\right.
\end{equation*}

\item \textit{Almost-orthogonality.} For every two lines $L_{1}\neq
L_{2}\subset V$, and every $f_{L_{1}}\in \mathcal{B}_{L_{1}}$, $f_{L_{2}}\in 
\mathcal{B}_{L_{2}},$ we have%
\begin{equation*}
\left\vert \mathcal{M}(f_{L_{1}},f_{L_{2}})[\tau ,\omega ]\right\vert =\frac{%
1}{\sqrt{N}},\text{ \ }
\end{equation*}%
for every $(\tau ,\omega )\in V.\medskip $
\end{enumerate}

\begin{figure}[h!] 
\includegraphics[clip,height=5cm]{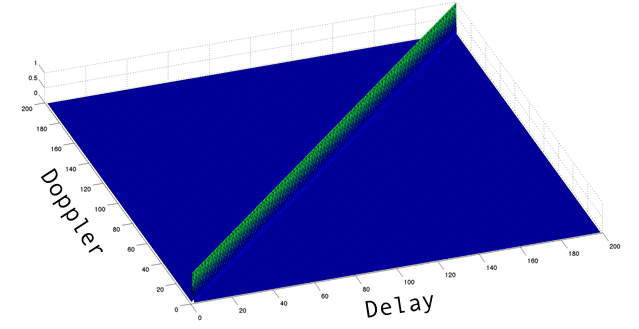}\\
\caption{$\left\vert 
\mathcal{M}(f_{L},f_{L})\right\vert $ for $L=\{(\protect\tau ,\protect\tau %
))\}.$}
\label{diagonal}
\end{figure}

Figure \ref{diagonal} demonstrates Property 1 of Theorem \ref{HT} for the
diagonal line.


\subsection{\textbf{The Weil (Spikes) System\label{WS}}}

\textit{\ }The group $G=SL_{2}(%
\mathbb{Z}
_{N})$ is non-commutative, but contains a special class of maximal
commutative subgroups called tori\footnote{%
There are order of $N^{2}$ tori in $SL_{2}(%
\mathbb{Z}
_{N}).$} \cite{GHS1, GHS2, B}. Each torus $T\subset G$ acts via the Weil
operators 
\begin{equation}
\rho (g):\mathcal{H\rightarrow \mathcal{H}}\text{, \ }g\in T.  \label{W}
\end{equation}%
This is a commutative collection of diagonalizable operators, and it admits 
\cite{GHS1, GHS2} a natural orthonormal basis $\mathcal{B}_{T}$ for $%
\mathcal{H}$, consisting of common eigensequences for all the operators (\ref%
{W})%
\begin{equation}
\left\{ 
\begin{array}{c}
\mathcal{B}_{T\text{ }}=\{\varphi _{T_{\chi }}\}; \\ 
\rho (g)\varphi _{T_{\chi }}=\chi (g)\varphi _{T_{\chi }},\text{ \ }g\in T,%
\text{ }%
\end{array}%
\right.  \label{BT}
\end{equation}%
where $\chi $ runs over \textit{characters} of $T$, i.e., functions $\chi
:T\rightarrow 
\mathbb{C}
^{\ast }$ with $\chi (g\cdot g^{\prime })=\chi (g)\chi (g^{\prime }),$ for
every $g,g^{\prime }\in T.$ \medskip

\begin{remark}
There is a small abuse of notation in \ref{BT}. There are two types of tori
in $G$, split tori and non-split tori \cite{GHS1, GHS2}. In each case, the
torus $T$ admits a unique non-trivial character $\chi _{q}$ of $T$---called
the \textit{quadratic character}---which takes the values $\chi _{q}(g)\in
\{\pm 1\},$ $g\in T.$ The dimension of the space $\mathcal{H}_{\chi _{q}}$
of sequences $\varphi _{T_{\chi _{q}}}$, which satisfy $\rho (g)\varphi
_{T_{\chi _{q}}}=\chi _{q}(g)\varphi _{T_{\chi _{q}}}$ is equal to $2$ or $0$%
, if $T$ is a split or non-split torus, respectively \cite{GHS1, GHS2}%
.\medskip
\end{remark}

Let us denote by 
\begin{equation*}
\mathcal{S}_{T}=\mathcal{B}_{T}\smallsetminus \mathcal{H}_{\chi _{q}},
\end{equation*}%
the set of sequences in $\mathcal{B}_{T},$ which are not associated with the
quadratic character. The system of all such sets $\mathcal{S}_{T},$ where $T$
runs over all tori in $G,$ will be called the \textit{Weil (spikes) system}.
We use the following result \cite{GHS1, GHS2}:\smallskip \medskip

\begin{theorem}
\label{WPeak}The Weil system satisfies the following properties:

\begin{enumerate}
\item \textit{Spike. }For every torus $T\subset G$, and every $\varphi
_{T}\in \mathcal{S}_{T}$, we have%
\begin{equation*}
\mathcal{M}(\varphi _{T},\varphi _{T})[\tau ,\omega ]=\left\{ 
\begin{array}{c}
1,\text{ if }\left( \tau ,\omega \right) =(0,0);\text{ \ \ \ \ \ \ \ \ \ \ }
\\ 
\leq \frac{2}{\sqrt{N}},\text{ in }\left\vert \cdot \right\vert \text{ if }%
\left( \tau ,\omega \right) \neq (0,0).%
\end{array}%
\right.
\end{equation*}

\item \textit{Almost-orthogonality.} For every two tori $T_{1},$ $%
T_{2}\subset G$, and every $\varphi _{T_{1}}\in \mathcal{S}_{T_{1}}$, $%
\varphi _{T_{2}}\in \mathcal{S}_{T_{2}},$ with $\varphi _{T_{1}}\neq \varphi
_{T_{2}},$ we have%
\begin{equation*}
\left\vert \mathcal{M}(\varphi _{T_{1}},\varphi _{T_{2}})[\tau ,\omega
]\right\vert \leq \left\{ 
\begin{array}{c}
\frac{4}{\sqrt{N}},\text{ if }T_{1}\neq T_{2}; \\ 
\frac{2}{\sqrt{N}},\text{ if }T_{1}=T_{2},%
\end{array}%
\right. \text{ \ }
\end{equation*}%
for every $(\tau ,\omega )\in V.\medskip $
\end{enumerate}
\end{theorem}

\begin{figure}[h!] 
\includegraphics[clip,height=5cm]{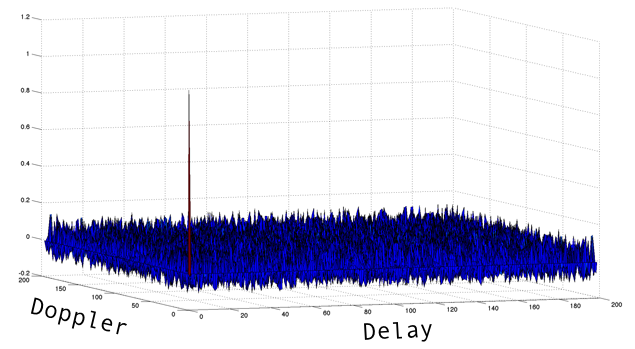}\\
\caption{$\mathcal{M}[\protect\varphi _{T},%
\protect\varphi _{T}]$ for $T=\{%
\protect\begin{pmatrix}
a & 0\protect \\ 
0 & a^{-1}%
\protect\end{pmatrix}%
;0\neq a\in 
\mathbb{Z}
_{N}\}.$}
\label{weil}
\end{figure}

Figure \ref{weil} illustrates Property 1 of Theorem \ref{WPeak}, applied to
the commutative subgroup of diagonal matrices in $G.$


\subsection{\textbf{The Heisenberg--Weil System}}

We define the \textit{Heisenberg--Weil system} of sequences. This is the
collection of sequences in $\mathcal{H}$, which are of the form $%
S_{L}=f_{L}+\varphi _{T}$, where $f_{L}$ and $\varphi _{T}$ are Heisenberg
and Weil sequences, respectively. The main technical result of this paper
is\smallskip :\medskip

\begin{theorem}
\label{HW}The Heisenberg--Weil system satisfies the properties\medskip

\begin{enumerate}
\item \textit{Flag. }For every line $L\subset V$, torus $T\subset G$, and
every flag $S_{L}=f_{L}+\varphi _{T},$ with $f_{L}\in \mathcal{B}_{L},$ $%
\varphi _{T}\in \mathcal{S}_{T}$, we have 
\begin{equation*}
\mathcal{M}(S_{L},S_{L})[\tau ,\omega ]=\left\{ 
\begin{array}{c}
2+\epsilon _{N}\text{,\ \ \ if }(\tau ,\omega )=(0,0);\text{ \ \ \ \ \ \ \ \
\ \ \ \ \ } \\ 
1+\varepsilon _{N}\text{, \ \ in }\left\vert \cdot \right\vert \text{ if }%
(\tau ,\omega )\in L\smallsetminus (0,0);\text{ } \\ 
\text{ }\varepsilon _{N}\text{,\ \ \ \ \ \ \ \ if }(\tau ,\omega )\in
V\smallsetminus L,\text{\ \ \ \ \ \ \ \ \ \ \ \ \ }%
\end{array}%
\right.
\end{equation*}%
where $|\epsilon _{N}|\leq \frac{4}{\sqrt{N}},$ and $|\varepsilon _{N}|\leq 
\frac{6}{\sqrt{N}}.\medskip $

\item \textit{Almost-orthogonality.} For every two lines $L_{1}\neq
L_{2}\subset V$, tori $T_{1},$ $T_{2}\subset G,$ and every two flags $%
S_{L_{j}}=f_{L_{j}}+\varphi _{T_{j}},$ with $f_{L_{j}}\in \mathcal{B}%
_{L_{j}} $, $\varphi _{T_{j}}\in \mathcal{S}_{T_{j}}$, $j=1,2,$ $\varphi
_{T_{1}}\neq \varphi _{T_{2}},$ we have%
\begin{equation*}
\left\vert \mathcal{M}(S_{L_{1}},S_{L_{2}})[\tau ,\omega ]\right\vert \leq
\left\{ 
\begin{array}{c}
\frac{9}{\sqrt{N}},\text{ if }T_{1}\neq T_{2}; \\ 
\frac{7}{\sqrt{N}},\text{ if }T_{1}=T_{2},%
\end{array}%
\right. \text{ \ }
\end{equation*}%
for every $(\tau ,\omega )\in V.\medskip $
\end{enumerate}
\end{theorem}

For a proof of Theorem \ref{HW} see Subsection \ref{P-HW}.\medskip

\begin{figure}[h!] 
\includegraphics[clip,height=5cm]{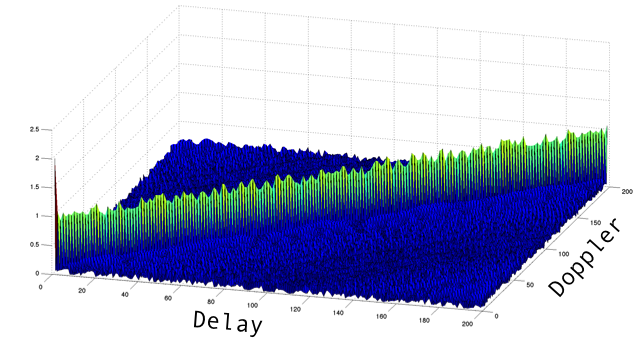}\\
\caption{$\left\vert \mathcal{M}[S_{L},S_{L}]\right\vert $ for Heisenberg--Weil flag with $L=\{(\protect\tau %
,\protect\tau )\}.$}
\label{hwf}
\end{figure}

Figure \ref{hwf} demonstrates Property 1 of Theorem \ref{HW} applied to the
diagonal line $L$ and a torus $T$ as in Figure \ref{weil}.


\begin{remark}
As a consequence of Theorem \ref{HW} we obtain families of $N+1$
almost-orthogonal flag sequences which can be used for solving the TFS
problem in $O(N\log N)$ operations, and GPS, channel estimation, radar, and
mobile communication problems in $O(m\cdot N\log N)$ operations for channel
of sparsity $m$.\medskip
\end{remark}

This completes our design of the Heisenberg--Weil flag sequences.

\section{\textbf{Formulas for Heisenberg--Weil Sequences\label{H--W For}}}

In order to implement the flag method it is important to have explicit
formulas for the Heisenberg and Weil sequences, which in particular enable
one to generate them with a low complexity procedure. In this section we
supply such effective description for all Heisenberg sequences, and for Weil
sequences associated with split tori

\subsection{\textbf{Formulas for Heisenberg Sequences}}

First we parametrize the lines in the time-frequency plane, and then we
provide explicit formulas for the orthonormal bases of sequences associated
with the lines.\medskip

\subsubsection{\textbf{Parametrization of Lines}}

The $N+1$ lines in the time-frequency plane $V=%
\mathbb{Z}
_{N}\times 
\mathbb{Z}
_{N}$ can be described in terms of their slopes. We have\medskip

\begin{itemize}
\item \textit{Lines with finite slope. }These are the lines of the form $%
L_{c}=$\textrm{span}$\{(1,c)\},$ $c\in 
\mathbb{Z}
_{N}$.\smallskip

\item \textit{Line with infinite slope. }This is the line $L_{\infty }=$%
\textrm{span}$\{(0,1)\}.\medskip \medskip $
\end{itemize}

\subsubsection{\textbf{Formulas}}

Using the above parametrization, we obtain\medskip

\begin{itemize}
\item \textit{Formulas for Heisenberg sequences associated with lines of
finite slope.}\textbf{\ }For $c\in 
\mathbb{Z}
_{N}$ we have the orthonormal basis 
\begin{equation*}
\mathcal{B}_{L_{c}}=\{f_{c,b}[n]=\frac{1}{\sqrt{N}}e^{\frac{2\pi i}{N}%
(-2^{-1}cn^{2}+bn)}\text{ ; \ }b\in 
\mathbb{Z}
_{N}\},
\end{equation*}%
of Heisenberg sequences associated with the line $L_{c}.\medskip $

\item \textit{Formulas for Heisenberg sequences associated with the line of
infinite slope. }We have the orthonormal basis \textit{\ }%
\begin{equation*}
\mathcal{B}_{L_{\infty }}=\{\delta _{b};\ b\in 
\mathbb{Z}
_{N}\},
\end{equation*}%
of Heisenberg sequences associated with the line $L_{\infty },$ where the $%
\delta _{b}$'s denote the Dirac delta functions.
\end{itemize}

\subsection{\textbf{Formulas for the Weil Sequences}}

We describe explicit formulas for the Weil sequences associated with split
tori \cite{GG, GHS1, GHS2}. First we parametrize the split tori in $G=SL_{2}(%
\mathbb{Z}
_{N})$, and then we write the explicit expressions for the orthonormal bases
of sequences associated with these tori.\medskip

\subsubsection{\textbf{Parametrization of Split Tori}}

A commutative subgroup $T\subset G$ is called \textit{split torus} \cite{B}
if for some $g\in G$ it is of the form $T=T_{g},$ with 
\begin{equation*}
T_{g}=g\cdot A\cdot g^{-1},
\end{equation*}%
where $A\subset G$ is the subgroup of all diagonal matrices, also called the 
\textit{standard} torus, i.e., 
\begin{equation*}
A=\left\{ 
\begin{pmatrix}
a & 0 \\ 
0 & a^{-1}%
\end{pmatrix}%
;\text{ \ }0\neq a\in 
\mathbb{Z}
_{N}\right\} .
\end{equation*}%
We denote by $\mathcal{T=\{}T_{g};$ $g\in G\}$ the set of all split tori in $%
G.$ It is not hard to verify that the number of elements in $\mathcal{T}$ is 
$\frac{N(N+1)}{2}.$ A direct computation shows that the collection of all $%
T_{g}$'s with \ 
\begin{equation}
g=%
\begin{pmatrix}
1 & b \\ 
c & 1+bc%
\end{pmatrix}%
,\text{ \ }b,c\in 
\mathbb{Z}
_{N},  \label{Par}
\end{equation}%
exhausts the set $\mathcal{T}$. Moreover, in (\ref{Par}) the torus $T_{g}$
can be written also as $T_{g^{\prime }},$ for $g\neq g^{\prime },$ only if $%
b\neq 0$ and 
\begin{equation*}
g^{\prime }=%
\begin{pmatrix}
1 & b \\ 
c & 1+bc%
\end{pmatrix}%
\begin{pmatrix}
0 & -b \\ 
b^{-1} & 0%
\end{pmatrix}%
.
\end{equation*}%
\medskip

\subsubsection{\textbf{Formulas}}

In order to provide the explicit formulas we need to develop some basic
facts and notations from the theory of multiplicative characters \cite{Ar}.
Consider the group $%
\mathbb{Z}
_{N}^{\ast }$ of all non-zero elements in $%
\mathbb{Z}
_{N},$ with multiplication modulo $N.$ A basic fact about this group is that
it is\textit{\ cyclic, }i.e., there exists an element $r\in 
\mathbb{Z}
_{N}^{\ast }$\textit{\ }such that 
\begin{equation*}
\mathbb{Z}
_{N}^{\ast }=\{1,r,r^{2},...,r^{N-2}\}.
\end{equation*}%
A function $\mathcal{\chi }:%
\mathbb{Z}
_{N}^{\ast }\rightarrow 
\mathbb{C}
^{\ast }$ is called \textit{multiplicative character} if $\chi (x\cdot
y)=\chi (x)\cdot \chi (y)$ for every $x,y\in 
\mathbb{Z}
_{N}^{\ast }.$ A way to write formulas for such functions is the following.
Choose $\zeta \in 
\mathbb{C}
$ which satisfies $\zeta ^{N-1}=1,$ i.e., $\zeta \in \mu _{N-1}=\{e^{\frac{%
2\pi i}{N-1}k};$ $k=0,...,N-2\}$, and define a multiplicative character by 
\begin{equation*}
\chi _{\zeta }(r^{d})=\zeta ^{d},\text{ }d=0,1,...,N-2.
\end{equation*}%
Running over all the $N-1$ possible such $\zeta $'s, we obtain all the
multiplicative characters of $%
\mathbb{Z}
_{N}^{\ast }.$ We are ready to write, in terms of the parametrization (\ref%
{Par}), the concrete eigensequences associated with each of the tori (see
Subsection \ref{WS}). \ We obtain\medskip

\begin{itemize}
\item \textit{Formulas for Weil sequences associated with the diagonal
torus. }For the diagonal torus $A$ we have 
\begin{equation*}
\mathcal{S}_{A}=\left\{ \varphi _{\chi _{\zeta }};\text{ \ }-1\neq \zeta \in
\mu _{N-1}\right\} ,
\end{equation*}%
where $\varphi _{\chi _{\zeta }}\in \mathcal{H}$ is the sequence defined by 
\begin{equation}
\varphi _{\chi _{\zeta }}[n]=\left\{ 
\begin{array}{c}
\frac{1}{\sqrt{N-1}}\chi _{\zeta }[n]\text{ \ if \ }n\neq 0,\text{ \ \ \ }
\\ 
0\text{ \ \ \ \ \ \ \ \ \ if \ }n=0.%
\end{array}%
\right.  \label{WS-A}
\end{equation}%
\medskip

\item \textit{Formulas for Weil sequences associated with the torus }$%
T_{u_{c}},$ \textit{for unipotent} $u_{c}\in G$\textit{. }For the torus $%
T_{u_{c}}$ associated with the unipotent element 
\begin{equation*}
u_{c}=%
\begin{pmatrix}
1 & 0 \\ 
c & 1%
\end{pmatrix}%
,\text{ }c\in 
\mathbb{Z}
_{N},
\end{equation*}%
we have 
\begin{equation*}
\mathcal{S}_{T_{u_{c}}}=\left\{ \varphi _{\chi _{\zeta }^{u_{c}}};\text{ }%
-1\neq \zeta \in \mu _{N-1}\right\} ,
\end{equation*}%
where $\varphi _{\chi _{\zeta }^{u_{c}}}\in \mathcal{H}$ is the sequence
defined by 
\begin{equation}
\varphi _{\chi _{\zeta }^{u_{c}}}[n]=e^{\frac{2\pi i}{N}(-2^{-1}cn^{2})}%
\cdot \varphi _{\chi _{\zeta }}[n],  \label{WS-u}
\end{equation}%
for every $n\in 
\mathbb{Z}
_{N},$ and $\varphi _{\chi _{\zeta }}$ is the sequence given by (\ref{WS-A}%
).\medskip

\item \textit{Formulas for Weil sequences associated with the torus }$T_{g},$
\textit{for non-unipotent} $g\in G$\textit{. }For the torus $T_{g}$
associated with the element 
\begin{equation*}
g=%
\begin{pmatrix}
1 & b \\ 
c & 1+bc%
\end{pmatrix}%
,\text{ \ }b,c\in 
\mathbb{Z}
_{N},\text{ }b\neq 0,
\end{equation*}%
we have 
\begin{equation*}
\mathcal{S}_{T_{g}}=\left\{ \varphi _{\chi _{\zeta }^{g}}\text{ };\text{ }%
-1\neq \zeta \in \mu _{N-1}\right\} ,
\end{equation*}%
where $\varphi _{\chi _{\zeta }^{g}}\in \mathcal{H}$ is the sequence defined
by 
\begin{eqnarray}
\varphi _{\chi _{\zeta }^{g}}[n] &=&C_{b}\cdot \frac{e^{\frac{2\pi i}{N}%
(2^{-1}\frac{1+bc}{b}n^{2})}}{\sqrt{N}}\times  \label{WS-g} \\
&&\times \sum_{\omega \in 
\mathbb{Z}
_{N}}e^{\frac{2\pi i}{N}\omega \cdot n}\cdot \left[ e^{\frac{2\pi i}{N}%
(-2^{-1}b\omega ^{2})}\cdot \varphi _{\chi _{\zeta }}[b\omega ]\right] , 
\notag
\end{eqnarray}%
for every $n\in 
\mathbb{Z}
_{N},$ and $\varphi _{\chi _{\zeta }}$ the sequence given by (\ref{WS-A}), $%
C_{b}=i^{-\frac{N-1}{2}}\QOVERD( ) {b}{N},$ with $\QOVERD( ) {\cdot }{N}$
the Legendre symbol.\medskip
\end{itemize}

The validity of Formula (\ref{WS-A}) is immediate from Identity (\ref{Sc}).
For a verification of Formulas (\ref{WS-u}) and (\ref{WS-g}), see Subsection %
\ref{Ver-WS}.\medskip

\begin{remark}[\textbf{Complexity of Heisenberg-Weil sequences}]
For concrete applications it is important to have low arithmetic complexity
algorithm generating the sequences. Note that the sequence (\ref{WS-g}) can
be computed in $O(N\log N)$ operations using FFT. We conclude that the all
Heisenberg sequences, and all Weil sequences associated with split tori, and
in particular the associated flag sequence, can be computed in at most $%
O(N\log N)$ operations.
\end{remark}

\section{\textbf{Computing the Matched Filter on a Line\label{MF on Line}}}

\begin{figure}[h!] 
\includegraphics[clip,height=5cm]{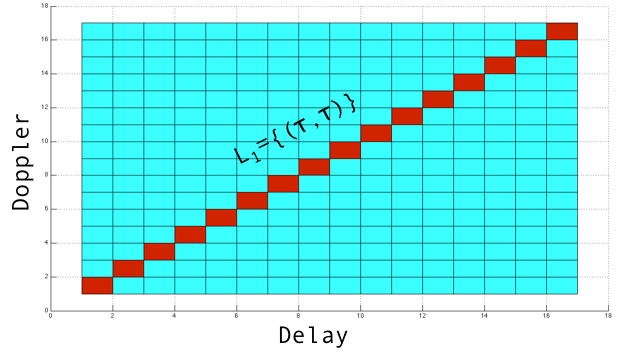}\\
\caption{$\mathcal{M}(\protect\varphi ,\protect\phi )[\protect\tau ,\protect\tau ]=%
\left[ \mathsf{m}_{\exp (2^{-1}n^{2})}\protect\varphi _{-}\ast \mathsf{m}%
_{\exp (-2^{-1}n^{2})}\overline{\protect\phi }\right] [\protect\tau ]$ on $%
L_{1}.$}
\label{M-on-L}
\end{figure}

Implementing the flag method, we need to compute in $O(N\log N)$ operations
the restriction of the MF matrix to any line in the time-frequency plane
(see Remark \ref{FFT}). In this section we provide algorithm that fulfills
this task. The upshot is---see Figure \ref{M-on-L} for illustration of the
case of the diagonal line---that the restriction of the MF matrix to a line
is a certain convolution that can be computed fast using FFT. Denote by $%
\mathcal{M(}\varphi ,\phi )[\tau ,\omega ]=\left\langle \varphi ,\text{ }\pi
(\tau ,\omega )\phi \right\rangle $ the matched filter associated with
sequences $\varphi ,\phi \in \mathcal{H}$, and by $\varphi \ast \phi \in 
\mathcal{H}$ their convolution%
\begin{equation}
\left( \varphi \ast \phi \right) [\tau ]=\sum_{n\in 
\mathbb{Z}
_{N}}\varphi _{-}[n]\cdot \phi _{\tau }[n],  \label{Con}
\end{equation}%
where $\varphi _{-}[n]=\varphi \lbrack -n]$, and $\phi _{\tau }[n]=\phi
\lbrack \tau +n],$ for every $\tau ,n\in 
\mathbb{Z}
_{N}.$

We consider two cases:\medskip

\begin{enumerate}
\item \textit{\ Formula on lines with finite slope and their shifts. }For $%
c\in 
\mathbb{Z}
_{N}$ consider the line $L_{c}=\{\tau \cdot (1,c)$ $;$ $\tau \in 
\mathbb{Z}
_{N}\},$ and for a fixed $\omega \in 
\mathbb{Z}
_{N}$ the shifted line $L_{c}^{\prime }=L_{c}+(0,\omega ).$ On $%
L_{c}^{\prime }$ we have\medskip\ 
\begin{eqnarray}
&&\mathcal{M(}\varphi ,\phi )[\tau \cdot (1,c)+(0,\omega )]  \label{M-on-Lc}
\\
&=&\left[ \mathsf{m}_{\exp (2^{-1}cn^{2}+\omega n)}\varphi _{-}\ast \mathsf{m%
}_{\exp (-2^{-1}cn^{2})}\overline{\phi }\right] [\tau ],  \notag
\end{eqnarray}%
\medskip where $\left[ \mathsf{m}_{\exp (2^{-1}cn^{2}+\omega n)}\varphi _{-}%
\right] [n]=e^{\frac{2\pi i}{N}(2^{-1}cn^{2}+\omega n)}\times \varphi
_{-}[n] $, $n\in 
\mathbb{Z}
_{N}$, and similar definition for the second expression, with $\overline{%
\phi }$ the complex conjugate of the sequence $\phi .\medskip $

\item \textit{Formula on the line with infinite slope and its shifts. }%
Consider the line $L_{\infty }=\{\omega \cdot (0,1)$ $;$ $\omega \in 
\mathbb{Z}
_{N}\},$ and for a fixed $\tau \in 
\mathbb{Z}
_{N}$ the shifted line $L_{\infty }^{\prime }=L_{\infty }+(\tau ,0).$ On $%
L_{\infty }^{\prime }$ we have\medskip\ 
\begin{equation}
\mathcal{M(}\varphi ,\phi )[\omega \cdot (0,1)+(\tau ,0)]=DFT(\varphi \cdot 
\overline{\phi _{\tau }})[\omega ],  \label{M-on-Lin}
\end{equation}%
for every $\omega \in 
\mathbb{Z}
_{N}.\medskip $
\end{enumerate}

The validity of Formula (\ref{M-on-Lin}) is immediate from the definition of
the matched filter. For a verification of Formula (\ref{M-on-Lc}) see
Subsection \ref{Just-M-on-Lc}.


\section{\textbf{Proofs\label{Proofs}}}

\subsection{\textbf{Proof of Theorem \protect\ref{HW} \label{P-HW}}}

\subsubsection{\textbf{Flag Property}}

Let $S_{L}=f_{L}+\varphi _{T}.$ We have%
\begin{eqnarray*}
\mathcal{M(}S_{L},S_{L}) &=&\mathcal{M(}f_{L},f_{L})+\mathcal{M(}%
f_{L},\varphi _{T}) \\
&&+\mathcal{M(}\varphi _{T},f_{L})+\mathcal{M(}\varphi _{T},\varphi _{T}).
\end{eqnarray*}%
We will show that%
\begin{equation}
\left\vert \mathcal{M(}\varphi _{T},f_{L})[\tau ,\omega ]\right\vert \leq 
\frac{2}{\sqrt{N}},\text{ \ }\tau ,\omega \in 
\mathbb{Z}
_{N}.  \label{B}
\end{equation}%
Noting that $\mathcal{M(}f_{L},\varphi _{T})[\tau ,\omega ]=\overline{%
\mathcal{M(}\varphi _{T},f_{L})[-\tau ,-\omega ]}$ we get from (\ref{B})
also the same bound for $\mathcal{M(}f_{L},\varphi _{T}).$ Having this,
using Theorems \ref{HT} and \ref{WPeak} we can deduce the Flag Property. So
assume $\pi (\ell )f_{L}=\psi (\ell )f_{L}$ for $\ell \in L.$ We proceed in
several steps:\smallskip

\textbf{Step 1.}\textit{\ For every }$v\in V$ \textit{we have} $\pi
(v)f_{L}\in \mathcal{B}_{L}.$ Indeed, for $\ell \in L$ we have 
\begin{eqnarray*}
\pi (\ell )[\pi (v)f_{L}] &=&e^{\frac{2\pi i}{N}(-\Omega (\ell ,v))}\pi
(v)\pi (\ell )f_{L} \\
&=&e^{\frac{2\pi i}{N}(-\Omega (\ell ,v))}\psi (\ell )[\pi (v)f_{L}],
\end{eqnarray*}%
where $\Omega :V\times V\rightarrow 
\mathbb{Z}
_{N}$ is the symplectic form $\Omega \lbrack (\tau ,\omega ),(\tau ^{\prime
},\omega ^{\prime })]=\tau \omega ^{\prime }-\omega \tau ^{\prime }.$
Namely, $\pi (v)f_{L}$ is eigensequence for $\pi (\ell )$ with character $%
\psi _{v}(\ell )=$ $e^{\frac{2\pi i}{N}(-\Omega (\ell ,v))}\psi (\ell ).$

By step 1, it is enough to bound the inner product%
\begin{equation}
\left\vert \left\langle \varphi _{T},f_{L}\right\rangle \right\vert \leq 
\frac{2}{\sqrt{N}}.  \label{E}
\end{equation}
\smallskip

\textbf{Step 2.}\textit{\ The bound (\ref{B}) holds for }$L_{\infty }.$
Indeed, then $f_{L_{\infty }}=\delta _{b}$ for some $b\in 
\mathbb{Z}
_{N},$ hence 
\begin{equation*}
\left\vert \left\langle \varphi _{T},f_{L_{\infty }}\right\rangle
\right\vert =|\varphi _{T}[b]|\leq \sup_{n\in 
\mathbb{Z}
_{N}}|\varphi _{T}[n]|.
\end{equation*}%
In \cite{GHS2} it was shown that for every Weil sequence $\varphi _{T}$ we
have 
\begin{equation*}
\sup_{n\in 
\mathbb{Z}
_{N}}|\varphi _{T}[n]|\leq \frac{2}{\sqrt{N}}.
\end{equation*}

\textbf{Step 3.}\textit{\ The bound (\ref{B}) holds for every line }$L.$ We
will use two lemmas. First, let $L,M\subset V$ be two lines, and $g\in G$
such that $gL=\{g\cdot \ell $ $;$ $\ell \in L\}=M.$ For a character $\psi
:L\rightarrow 
\mathbb{C}
^{\ast },$ define the character $\psi ^{g}:M\rightarrow 
\mathbb{C}
^{\ast },$ by $\psi ^{g}(g\cdot \ell )=\psi (\ell )$, for every $\ell \in L.$
We have\medskip\ 

\begin{lemma}
\label{Eigen-1}Suppose $f_{L}$ is a $\psi $-eigensequence for $L,$ i.e., $%
\pi (\ell )f_{L}=\psi (\ell )f_{L},$ for every $\ell \in L.$ Then the
sequence $f_{M}=\rho (g)f_{L}$ is $\psi ^{g}$-eigensequence for $M.\medskip
\medskip $
\end{lemma}

For a proof of Lemma \ref{Eigen-1}, see Subsection \ref{PEigen-1}.\medskip

For the second lemma, consider a torus $T\subset G,$ and an element $g\in G.$
Then we can define a new torus $T_{g}=gTg^{-1}=\{g\cdot h\cdot g^{-1};$ \ $%
h\in T\}.$ For a character $\chi :T\rightarrow 
\mathbb{C}
^{\ast },$ we can associate a character $\chi ^{g}:T_{g}\rightarrow 
\mathbb{C}
^{\ast }$, by $\chi ^{g}(g\cdot h\cdot g^{-1})=\chi (h),$ for every $h\in T.$
We have\medskip\ 

\begin{lemma}
\label{Eigen-2}Suppose $\varphi _{T}$ is a $\chi $-eigensequence for $T,$
i.e., $\rho (h)\varphi _{T}=\chi (h)\varphi _{T},$ for every $h\in T.$ Then
the sequence $\varphi _{T_{g}}=\rho (g)\varphi _{T}$ is $\chi ^{g}$%
-eigensequence for $T_{g}.\medskip \medskip $
\end{lemma}

For a proof of Lemma \ref{Eigen-2}, see Subsection \ref{PEigen-2}.\medskip
\medskip

Now we can verify Step 3. Indeed, given a line $L\subset V,$ there exists $%
g\in G$ such that $g\cdot L=L_{\infty }.$ In particular, by Lemma \ref%
{Eigen-1} we get that $f_{L_{\infty }}=\rho (g)f_{L}$ is up to a unitary
scalar in $\mathcal{B}_{L_{\infty }}.$ In addition, by Lemma \ref{Eigen-2}
we know that $\varphi _{T_{g}}=\rho (g)\varphi _{T}$ is up to a unitary
scalar in $\mathcal{B}_{T_{g}}.$ Finally, we have 
\begin{eqnarray*}
\left\langle \varphi _{T},f_{L}\right\rangle &=&\left\langle \rho (g)\varphi
_{T},\rho (g)f_{L}\right\rangle \\
&=&\left\langle \varphi _{Tg},f_{L_{\infty }}\right\rangle ,
\end{eqnarray*}%
where the first equality is by the unitarity of $\rho (g)$. Hence, by Step
2, we get the desired bound also in this case. \medskip

\paragraph{\textbf{Proof of Lemma \protect\ref{Eigen-1} \label{PEigen-1}}}

For $\ell \in L$ we have 
\begin{eqnarray*}
\pi (g\cdot \ell )f_{M} &=&\pi (g\cdot \ell )\rho (g)f_{L} \\
&=&\rho (g)\pi (\ell )f_{L} \\
&=&\psi (\ell )\rho (g)f_{L} \\
&=&\psi ^{g}(g\cdot \ell )f_{M},
\end{eqnarray*}%
where the second equality is by Identity (\ref{S})$.$ This completes the
proof of Lemma \ref{Eigen-1}.

\paragraph{\textbf{Proof of Lemma \protect\ref{Eigen-2} \label{PEigen-2}}}

\bigskip For $h\in T$ we have 
\begin{eqnarray*}
\rho (g\cdot h\cdot g^{-1})\varphi _{T_{g}} &=&\rho (g\cdot h\cdot
g^{-1})\rho (g)\varphi _{T} \\
&=&\rho (g)\rho (h)\varphi _{T} \\
&=&\chi (h)\rho (g)\varphi _{T} \\
&=&\chi ^{g}(g\cdot h\cdot g^{-1})\varphi _{T_{g}},
\end{eqnarray*}%
where the second equality is because $\rho $ is homomorphism (see Theorem %
\ref{S-vN}). This completes our proof of Lemma \ref{Eigen-2}, and of the
Flag Property.\medskip

\subsubsection{\textbf{Almost Orthogonality }}

Let $S_{L_{j}}=f_{L_{j}}+\varphi _{Tj},$ $j=1,2,$ as in the assumptions. We
have%
\begin{eqnarray*}
\mathcal{M(}S_{L_{1}},S_{L_{2}}) &=&\mathcal{M(}f_{L_{1}},f_{L_{2}})+%
\mathcal{M(}f_{L_{1}},\varphi _{T_{2}}) \\
&&+\mathcal{M(}\varphi _{T_{1}},f_{L_{2}})+\mathcal{M(}\varphi
_{T_{1}},\varphi _{T_{2}}).
\end{eqnarray*}%
The result now follows from Theorem \ref{HT}, Theorem \ref{WPeak}, and the
bound (\ref{B}). This completes our proof of the Almost Orthogonality
Property, and of Theorem \ref{HW}.

\subsection{\textbf{Verification of Formulas (\protect\ref{WS-u}), and (%
\protect\ref{WS-g}) \label{Ver-WS}}}

\textbf{\ }The idea is to use the fact that for $g\in G,$ the explicit Weil
operator $\rho (g)$ maps the explicit set $\mathcal{S}_{A}$ to the set $%
\mathcal{S}_{T_{g}},$ $T_{g}=gAg^{-1}.$ In details, for a character $\chi
:A\rightarrow 
\mathbb{C}
^{\ast }$ and an element $g\in G$ define the character $\chi
^{g}:T_{g}\rightarrow 
\mathbb{C}
^{\ast },$ by $\chi ^{g}(g\cdot h\cdot g^{-1})=\chi (h)$, for every $h\in A.$
Using Lemma \ref{Eigen-2} we deduce that if $\varphi _{\chi }\in \mathcal{S}%
_{A}$ is eigensequence of $A$ with character $\chi $, then $\varphi _{\chi
^{g}}=\rho (g)\varphi _{\chi }\in \mathcal{S}_{T_{g}}$ is eigensequence of $%
T_{g}$ with character $\chi ^{g}.$ Specializing to the character $\chi =\chi
_{\zeta },$ $-1\neq \zeta \in \mu _{N-1},$ of $A,$ and the associated
sequence $\varphi _{\chi _{\zeta }}\in \mathcal{S}_{A}$ given by (\ref{WS-A}%
), we can proceed to verify the formulas.\medskip

\subsubsection{\textbf{Verification of Formula (\protect\ref{WS-u})}}

For the unipotent element%
\begin{equation*}
u_{c}=%
\begin{pmatrix}
1 & 0 \\ 
c & 1%
\end{pmatrix}%
,\text{ }c\in 
\mathbb{Z}
_{N},
\end{equation*}%
we have 
\begin{eqnarray*}
\varphi _{\chi _{\zeta }^{u_{c}}}[n] &=&\left[ \rho (u_{c})\varphi _{\chi
_{\zeta }}\right] [n] \\
&=&e^{\frac{2\pi i}{N}(-2^{-1}cn^{2})}\cdot \varphi _{\chi _{\zeta }}[n]%
\text{ },
\end{eqnarray*}%
where the second equality is by Formula (\ref{C}). This completes our
verification of Formula (\ref{WS-u}).\medskip

\subsubsection{\textbf{Verification of Formula (\protect\ref{WS-g})}}

For the element%
\begin{equation*}
g=%
\begin{pmatrix}
1 & b \\ 
c & 1+bc%
\end{pmatrix}%
,\text{ \ }b,c\in 
\mathbb{Z}
_{N},\text{ }b\neq 0,
\end{equation*}%
its Bruhat decomposition is%
\begin{equation}
\begin{pmatrix}
1 & b \\ 
c & 1+bc%
\end{pmatrix}%
=%
\begin{pmatrix}
1 & 0 \\ 
\frac{1+bc}{b} & 1%
\end{pmatrix}%
\begin{pmatrix}
\text{ \ }0 & 1 \\ 
-1 & \text{ }0%
\end{pmatrix}%
\begin{pmatrix}
1 & 0 \\ 
b & 1%
\end{pmatrix}%
\begin{pmatrix}
b^{-1} & 0 \\ 
0 & b%
\end{pmatrix}%
.  \label{BD}
\end{equation}%
This implies that for $n\in 
\mathbb{Z}
_{N}$ we have 
\begin{eqnarray*}
\varphi _{\chi _{\zeta }^{g}}[n] &=&\left[ \rho (g)\varphi _{\chi _{\zeta }}%
\right] [n] \\
&=&C_{b}\cdot \frac{e^{\frac{2\pi i}{N}(-2^{-1}\frac{1+bc}{b}n^{2})}}{\sqrt{N%
}}\times \\
&&\times \sum_{\omega \in 
\mathbb{Z}
_{N}}e^{\frac{2\pi i}{N}\omega \cdot n}\cdot \left[ e^{\frac{2\pi i}{N}%
(-2^{-1}b\omega ^{2})}\cdot \varphi _{\chi _{\zeta }}[b\omega ]\right] ,
\end{eqnarray*}%
with $C_{b}=i^{-\frac{N-1}{2}}\QOVERD( ) {b}{N},$ where in the second
equality we use identity (\ref{BD}), the fact that $\rho $ is homomorphism,
and the Formulas (\ref{F}), (\ref{C}), (\ref{Sc}). This completes our
verification of Formula (\ref{WS-g}).\medskip

\subsection{\textbf{Verification of Formula (\protect\ref{M-on-Lc}) \label%
{Just-M-on-Lc}}}

We verify Formula (\ref{M-on-Lc}) for the matched filter $\mathcal{M(}%
\varphi ,\phi ),$ $\varphi ,\phi \in \mathcal{H}$, restricted to a line with
finite slope. We proceed in two steps:\medskip

\textbf{Step 1.}\textit{\ The formula holds for the line }$L_{0}$\textit{\
and its shifts.} Indeed, for a fixed $\omega \in 
\mathbb{Z}
_{N}$\textit{\ }we compute the matched filter on $L_{0}^{\prime
}=L_{0}+(0,\omega )=\{(\tau ,\omega );$ $\tau \in 
\mathbb{Z}
_{N}\}.$ We get%
\begin{eqnarray*}
\mathcal{M(}\varphi ,\phi )[\tau ,\omega ] &=&\left\langle \varphi ,\text{ }%
\pi (\tau ,\omega )\phi \right\rangle \\
&=&\left\langle \varphi ,\text{ }e^{\frac{2\pi i}{N}\omega n}\cdot \phi
\lbrack n+\tau ]\right\rangle \\
&=&\left\langle \text{ }e^{-\frac{2\pi i}{N}\omega \cdot n}\cdot \varphi
\lbrack n],\text{ }\phi \lbrack n+\tau ]\right\rangle \\
&=&\left[ \mathsf{m}_{\exp (\omega n)}\varphi _{-}\ast \overline{\phi }%
\right] [\tau ],
\end{eqnarray*}%
where the fourth equality is by the definition (\ref{Con}) of $\ast ,$ and
the definition (\ref{M-on-Lc}) of $\mathsf{m}_{\exp (\cdot )}.\medskip $

\textbf{Step 2. }\textit{The formula holds for the lines }$L_{c},$\textit{\ }%
$c\in 
\mathbb{Z}
_{N},$ \textit{and their shifts. }Indeed, the element 
\begin{equation*}
u_{-c}=%
\begin{pmatrix}
1 & 0 \\ 
-c & 1%
\end{pmatrix}%
\in G,
\end{equation*}%
satisfies 
\begin{equation}
\left\{ 
\begin{array}{c}
u_{-c}\cdot (1,c)=(1,0), \\ 
u_{-c}\cdot (0,\omega )=(0,\omega ).%
\end{array}%
\right.  \label{u-c}
\end{equation}%
For a fixed $\omega \in 
\mathbb{Z}
_{N}$\textit{\ }we compute the matched filter on $L_{c}^{\prime
}=L_{c}+(0,\omega )=\{\tau \cdot (1,c)+(0,\omega );$ $\tau \in 
\mathbb{Z}
_{N}\}.$ We get 
\begin{eqnarray*}
&&\mathcal{M(}\varphi ,\phi )[\tau \cdot (1,c)+(0,\omega )] \\
&=&\left\langle \varphi ,\text{ }\pi \lbrack \tau \cdot (1,c)+(0,\omega
)]\phi \right\rangle \\
&=&\left\langle \rho (u_{-c})\varphi ,\text{ }\rho (u_{-c})\pi \lbrack \tau
\cdot (1,c)+(0,\omega )]\phi \right\rangle \\
&=&\left\langle \rho (u_{-c})\varphi ,\text{ }\pi (\tau ,\omega )\rho
(u_{-c})\phi \right\rangle \\
&=&\mathcal{M(}\rho (u_{-c})\varphi ,\rho (u_{-c})\phi )[\tau ,\omega ] \\
&=&\left[ \mathsf{m}_{\exp (2^{-1}cn^{2}+\omega n)}\varphi _{-}\ast \mathsf{m%
}_{\exp (-2^{-1}cn^{2})}\overline{\phi }\right] [\tau ],
\end{eqnarray*}%
where, the second equality is by the unitarity of $\rho ,$ the third
equality is by Identities (\ref{S}), (\ref{u-c}), and the last equality is
by Formula (\ref{C}) and Step 1 above.

This confirms Step 2, and completes our verification of Formula (\ref%
{M-on-Lc}).

\end{document}